\documentclass[aip,apl,%
 amsmath,amssymb,
 reprint
]{revtex4-1}
\usepackage{color}
\usepackage{hyperref}
\hypersetup{
     colorlinks   = true,
     citecolor    = blue,
     linkcolor    = blue
           }
\usepackage{graphicx}
\usepackage{dcolumn}
\usepackage{bm}

\begin{document}

\preprint{AIP/123-QED}

\title[]{Magnetic anisotropy in Permalloy: hidden quantum mechanical features}

\author{Debora C M Rodrigues}
\affiliation{Department of Physics and Astronomy, Uppsala University, Box 516, SE-751 20 Uppsala, Sweden}
\affiliation{Faculdade de F\'isica, Universidade Federal do Par\'a, CEP 66075-110, Bel\' em-PA, Brazil}
\author{Angela B Klautau}
\affiliation{Faculdade de F\'isica, Universidade Federal do Par\'a, CEP 66075-110, Bel\' em-PA, Brazil}
\author{Alexander Edstr\"om}
\author{Jan Rusz}
\author{Lars Nordstr\"om}
\author{Manuel Pereiro}
\author{Bj\"orgvin Hj\"orvarsson}
\author{Olle Eriksson}
\affiliation{Department of Physics and Astronomy, Uppsala University, Box 516, SE-751 20 Uppsala, Sweden}

\date{\today}

\begin{abstract}
By means of relativistic, first principles calculations, we investigate the microscopic origin of the vanishingly low magnetic anisotropy of Permalloy, here proposed to be intrinsically related to the local symmetries of the alloy. It is shown that the local magnetic anisotropy of individual atoms in Permalloy can be several orders of magnitude larger than that of the bulk sample, and 5-10 times larger than that of elemental Fe or Ni. We, furthermore, show that locally there are several easy axis directions that are favored, depending on local composition. The results are discussed in the context of perturbation theory, applying the relation between magnetic anisotropy and orbital moment. Permalloy keeps its strong ferromagnetic nature due to the exchange energy to be larger than the magnetocrystalline anisotropy. Our results shine light on the magnetic anisotropy of permalloy and of magnetic materials in general, and in addition enhance the understanding of pump-probe measurements and ultrafast magnetization dynamics. 
\end{abstract}


\maketitle

%
\textit{Introduction} -- Random alloys can be viewed as a distribution of clusters of different composition, that have an underlying crystal structure in common. The configurational space is enormous for these systems and any macroscopic property is the result of averaging of a immense amount of local clusters with different configuration and composition~\cite{abrikosov}. Random alloys often have properties that stand out from the pure elements they are build up from, i.e. the mixing of elements may produce properties that are completely unexpected.  

One of the most prominent examples is Permalloy (Py), the common name for Fe$_x$Ni$_{1-x}$ alloys with $x \sim 0.2$  and fcc crystal structure. These alloys are characterized by strong ferromagnetism, high permeability, vanishingly low magnetic anisotropy energy (MAE), and low damping parameter~\cite{coey}. These attributes elevate Py to a standard material in magnetism and    advantageous soft magnet for technological applications. 
 
One might ask what the mechanism of the vanishing MAE of Py really is. One attempt to explain it is the resultant MAE picture~\cite{mckeehan}, which reinforces that an appropriate mixture of two elements with distinct easy axis (as bcc Fe and fcc Ni with $\langle 001 \rangle$ and $\langle 111 \rangle$ direction, respectively) would result in a material without strong preferential easy magnetization axis, i.e. a low MAE. Nevertheless, recent experiments suggest that the Fe-Ni hybridization in the alloy environment is the major cause of low MAE in Py, rather than the easy axis of its separated constituents.~\cite{Yin}

Other experiments show the existence of orbital moments at the individual chemical species in Py~\cite{XMCD}. Additionally, it is known that the magnetic anisotropy is proportional to the anisotropy of orbital moment for transition metals~\cite{bruno,skomski}. Thus, the anisotropy at atomic level may exist, although diminished at the bulk. 
Considering this, as a random alloy, Py may be viewed as a huge ensemble of interconnected clusters of Fe-Ni atoms distributed on an fcc lattice, in which the macroscopic properties reflect the configurational average of different such clusters. Then different parts of the alloy would have competing local anisotropies, that effectively average out, leading to a fairly isotropic state. This microscopic scenario has, to the best of our knowledge, not been considered so far as a possible mechanism for low MAE in Py. 

For that, in this Letter, we present first principles calculations to investigate this local MAE. However it is not easy to directly study the MAE, particularly not from first principles theory as it is hard to uniquely and accurately decompose the energy into local contributions and the numerical challenges are countless. Instead we will study another related quantity induced by the spin-orbit coupling (SOC), namely the local anisotropy of the orbital moments, since it is known to reveal also information about the MAE~\cite{bruno}. With this information we investigate the role of these local competing anisotropies and how they reveal information about the soft magnetic behavior of Py.


\textit{Method} -- The study was designed as the following: first, we performed ab-initio calculations of a fcc matrix ($\sim$~12500 atoms) of a Virtual Crystal Approximation (VCA) medium of Py (Py-VCA) and lattice parameter of 3.54~\AA~\cite{XMCD}. The fcc unit cell was considered to have the same number of valence electrons as Py (9.6~$ \mathrm{e^-}$). After the self-consistent procedure, clusters composed by Fe and Ni, with different configurations, were embedded in the Py-VCA matrix. The cluster region was self-consistently updated while the potential parameters of the Py-VCA matrix were kept fixed. Finally, the magnetic spin and orbital moments were computed for every site in the cluster.

\begin{figure}[ht]
	\begin{center}
     \includegraphics[width=0.8\linewidth]{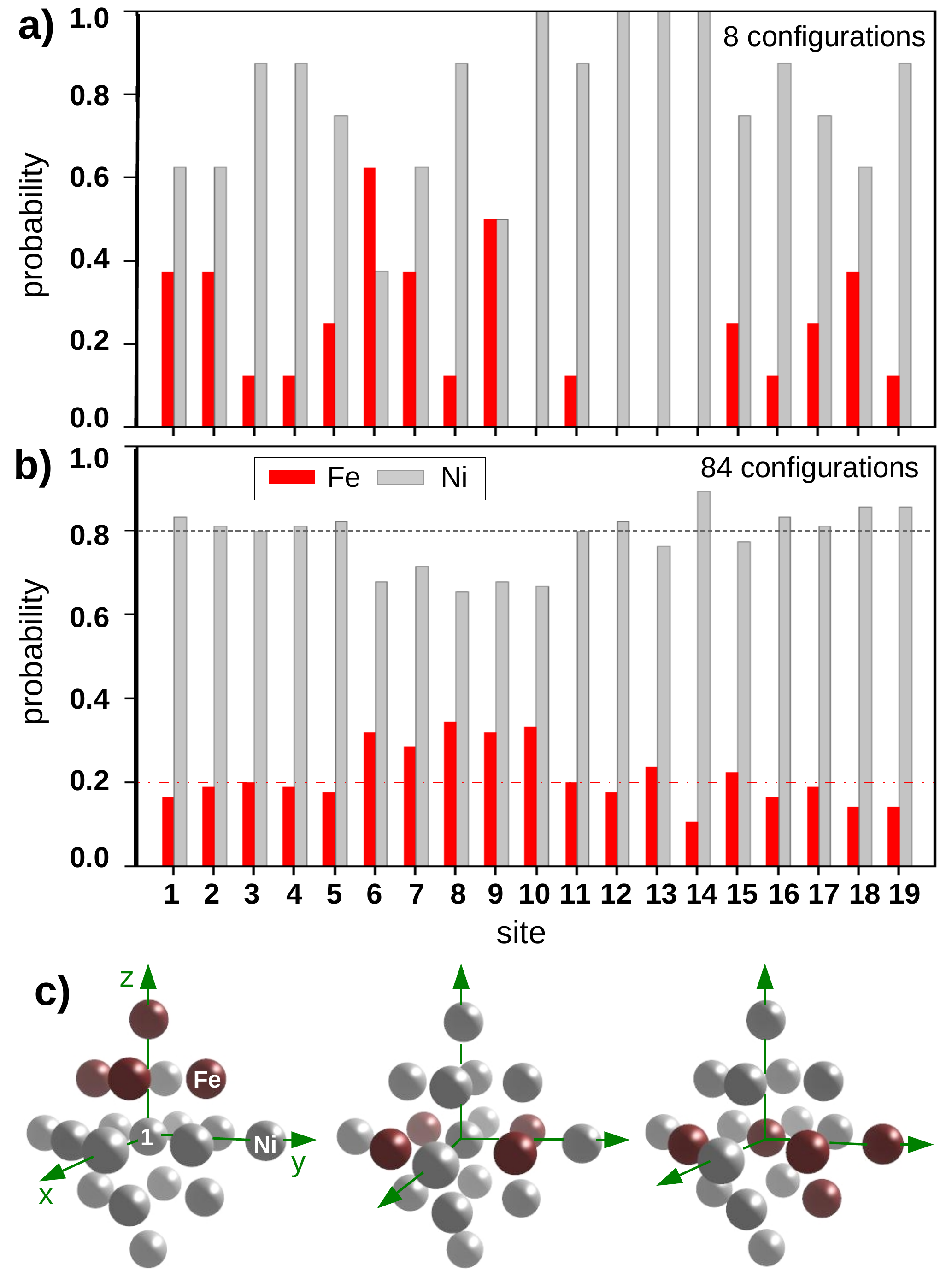}
	 \caption{(Color online) Distributions of Fe (red) and Ni (grey) atoms at cluster sites (1:~central site, 2-13:~first neigbours and 14-19:~second neigbours), considering (a) 8 or (b) 84 configurations. Three examples of configurations that may be found in Permalloy are illustrated in (c). 
	 The dashed (dot-dashed) line represent the average Fe (Ni) concentration.}\label{fig1}
	\end{center}
\end{figure}

As the configuration space for the clusters is vast, our investigation does not cover all possible configurations. 
We have, however, investigated a large number of geometries (84 excluding configurations that are symmetrical to these), and we illustrate as an example a few typical geometries in Fig.~\ref{fig1}c. These clusters present distinct configurations due to the Fe distribution. Note that, locally in a cluster, the number of Fe and Ni atoms can vary, although a configurational average over all clusters of the material would naturally result in a concentration of Fe and Ni that reflected the alloy concentration, i.e. 20 \% Fe and 80 \% Ni (see dashed lines in Fig.~\ref{fig1}b). 

In our investigations we choose to study clusters with 19 atoms, of which 4 or 5 are Fe atoms and the rest being Ni atoms. The atoms are sorted from a central site (labeled 1), followed by its first (labeled 2-13 ) and second neighbors (labeled 14-19 ).

The electronic structure and magnetism of VCA-Py and the clusters were evaluated using the first-principles real-space linear muffin-tin orbital method within the atomic sphere approximation (RS-LMTO-ASA)~\cite{Ref-RS1, Ref-RS2, rs-1, rs-2}. This method follows the steps of the LMTO-ASA formalism~\cite{Ref-And-SO} but uses the recursion method~\cite{Ref-Recorrencia2} to solve the eigenvalue problem directly in real space. The calculations presented here are fully self-consistent, the exchange and correlation terms were treated within the local spin density approximation (LSDA)~\cite{Ref-BH}, and the SOC term was included at each variational step~\cite{Ref-SOC1,Ref-SOC2}. The RS-LMTO-ASA method is particularly designed to treat low symmetry systems as the embedded clusters presented here, without the need of periodic boundary conditions.

Regarding the calculations for the matrix of VCA-Py, the resulting spin moment ($m_s$) is $1.12~\mu_B$ per atom, which is in acceptable agreement with the experimental value of approximately $1.0~\mu_B$ per atom~\cite{XMCD,Yin} and previous calculations~\cite{Yin,Yu,Minar}. Therefore, we conclude that the effective medium that is considered to host the different clusters reproduces the main features of Py. 

For the different clusters in this investigation we have estimated the local anisotropy from a well defined quantity -- the orbital moment anisotropy, which is the difference of the orbital moment projection $L$ for two different global quantization axes $\Delta L = L_{\mathbf{\hat n_1}} - L_{\mathbf{\hat n_2}}$. Since  $\Delta L$ is defined as a local quantity,  it is numerically easy to evaluate from first principles theory in contrast to the tiny energy difference needed for the MAE. It is established that the energy difference between two states with the magnetization direction along two different global directions is  $E_{\mathrm{MAE}} = - \frac{\xi}{4\mu_B} \Delta L$~\cite{bruno}, where $\xi$ is the SOC constant. One of the key assumptions in deriving this relation is that spin diagonal matrix elements of the spin orbit coupling should dominate the contribution to the MAE~\cite{candersson}. Since Py is a strong ferromagnet (the majority spin band is essentially filled) only minority spin states contribute to the density of states at the Fermi energy, and this criterion is expected to be fulfilled. When minority spin states dominate the MAE, the easy axis is parallel to the direction of maximum orbital magnetic moment. To exemplify the numerical advantage of the approach adopted here, we note that values of 1~$\mathrm{\mu}\mathrm{Ry}$ for the MAE are related to orbital anisotropies of $10^{-4}~\mathrm{\mu_B}$, which are values well defined by the method's precision. Thus, it serves well as the relevant quantity to evaluate and to quantify the local anisotropies in alloys.


\textit{Results} -- Before we discuss the results of the MAE, we note that for all configurations investigated here, the calculated individual moments were close to 
 $m^{Fe}_s = 2.30~\mathrm{\mu_B}$ and $m^{Ni}_s =0.61~\mathrm{\mu_B}$, for spin, and $L^{Fe} = 0.045~\mathrm{\mu_B}$ and $L^{Ni} = 0.031~\mathrm{\mu_B}$ for orbital ones. These values are in agreement with previous theoretical~\cite{Yu,Minar} and experimental~\cite{XMCD} studies.

\begin{figure}[ht]
	\begin{center}
\includegraphics[width=0.95\linewidth,  trim= 0.0cm 7.5cm 0.0cm 0.0cm, clip=true]{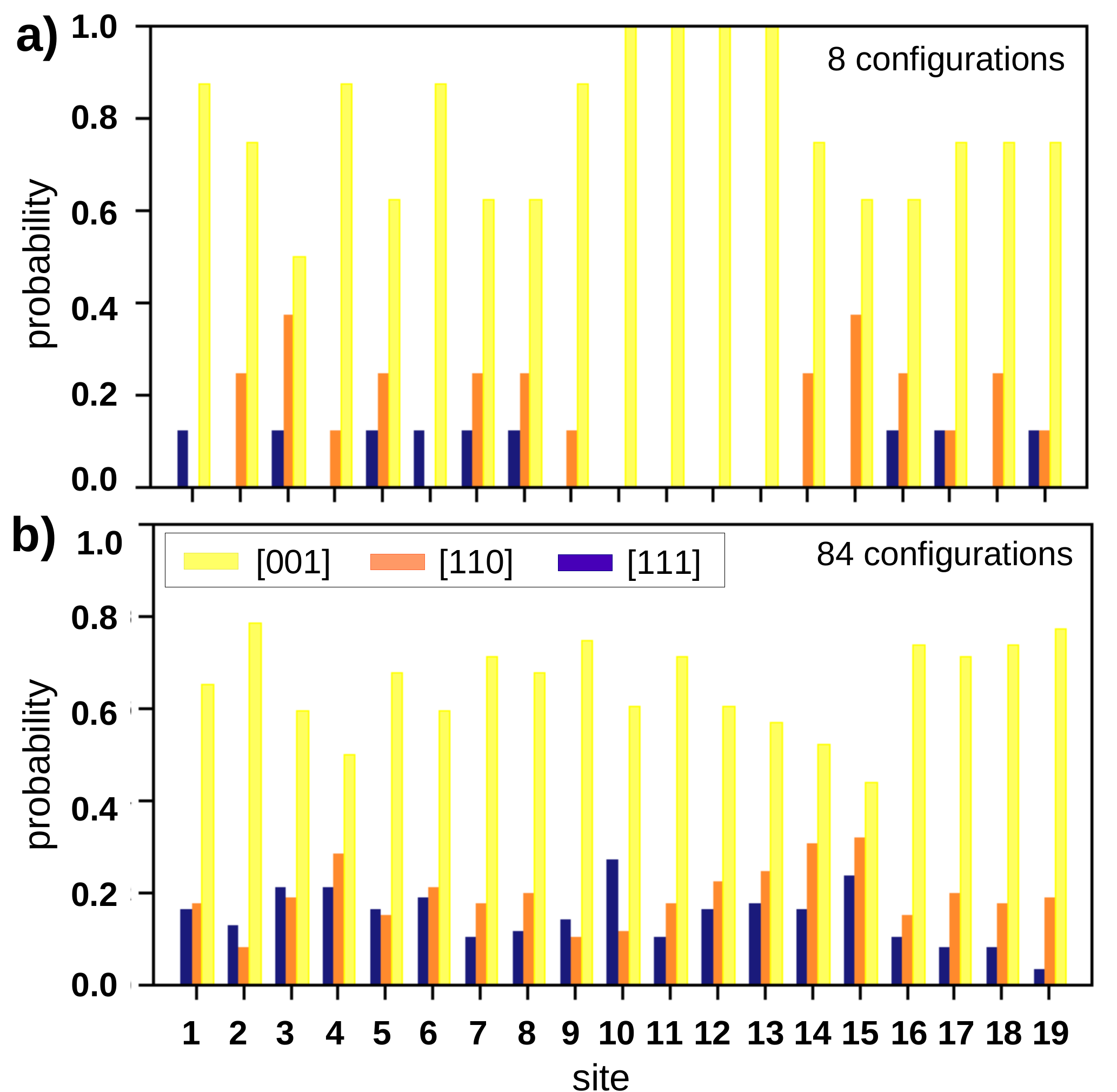} 
		\caption{(Color online) Likelyhood of different easy axis directions for each of the 19 atomic sites considered for each cluster. Averages are formed from 8 configurations (a) and 84 configurations (b). The easy axis direction are represented by the color bars.} \label{fig2-new}  
	\end{center}
\end{figure}

From each cluster of our investigation, we estimated the $\Delta L$, between two magnetization directions, for all atoms. Therefore, the global direction $\mathbf{\hat n_1}=[001]$ was considered as reference and the orbital moment anisotropy computed as $\Delta L = L_{[001]} - L_{\mathbf{\hat n_2}}$, with $\mathbf{\hat n_2}=[110]$ and $[111]$ directions. Comparing the $\Delta L$ values one can obtain the direction of maximum L, i.e. the local easy axis. This information is summarized in Fig.~\ref{fig2-new}, which shows the distributions of easy axis directions per site. In Fig.~\ref{fig2-new}a we show results formed from an average over 8 different clusters and in Fig.2b the average is made over 84 different configurations. We note that the number of configurations favoring the $[100]$ easy axis is larger than the $[110]$ and $[111]$ easy axis directions. We will return to this fact below. 

\begin{figure}[ht]
	\begin{center}
\includegraphics[width=\linewidth]{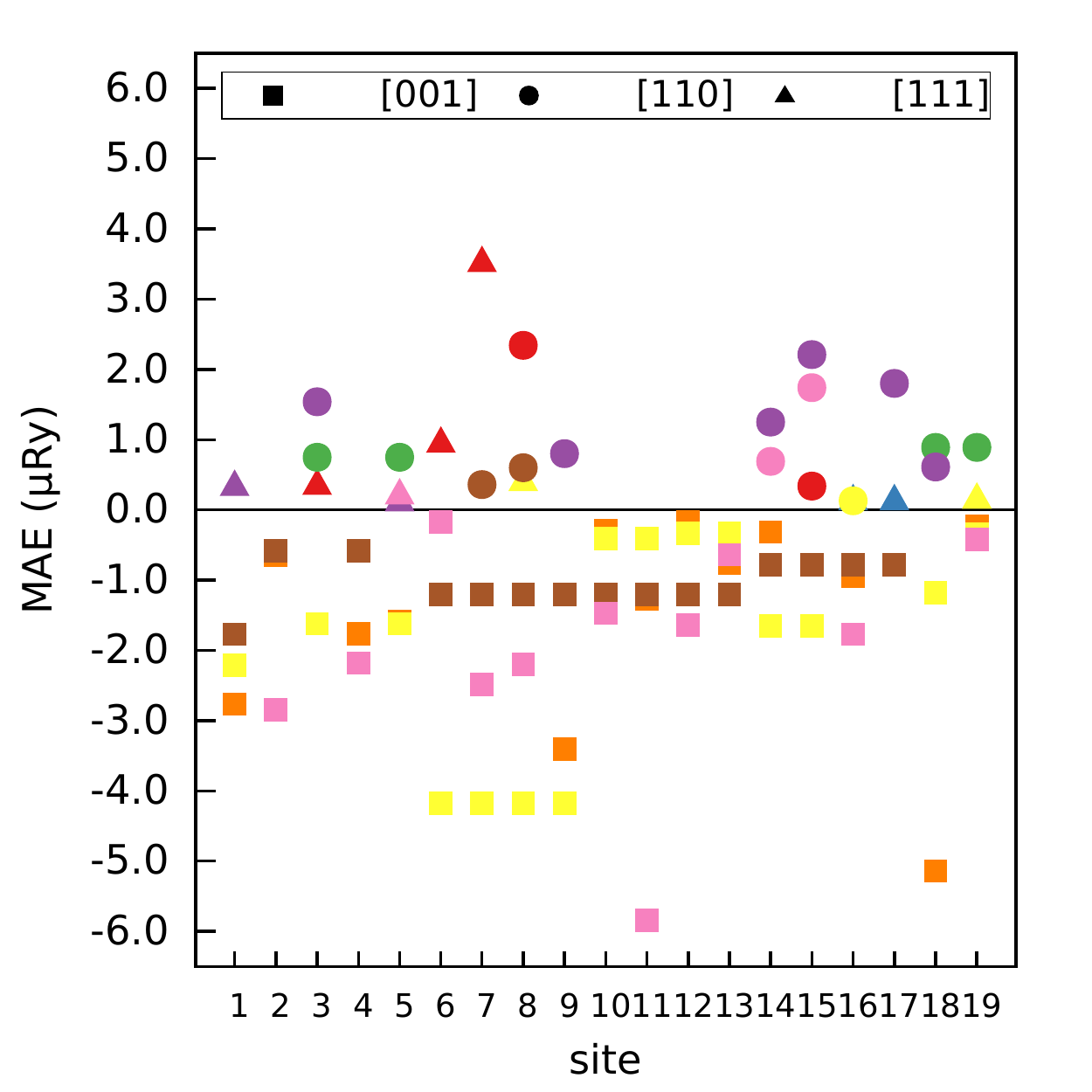} 
		\caption{(Color online) MAE per site with the easy axis direction represented by squares (for $[100]$), circles (for $[110]$) and triangles (for $[111]$ direction). Negative values of MAE symbolize an $[001]$ easy axis. The 8 different configurations considered here are represented by different colors. Note that some data points are superposed, since the same MAE value is found for an atom placed in a given site of different configurations.} \label{fig3}  
	\end{center}
\end{figure}

In addition to these values of $\Delta L$ we also estimated the site resolved E$_\mathrm{MAE}$, for each type of cluster. For that, we used the calculated SOC constants for $\xi_{\mathrm{Fe}} = 4.0~\mathrm{mRy}$ and $\xi_{\mathrm{Ni}} = 6.7~\mathrm{mRy}$. 
The values of E$_\mathrm{MAE}$ and the direction of the easy axis for each atom are shown in Fig.~\ref{fig3} (for sake of simplicity only 8 configurations are shown). Note from the figure that we show local easy axis directions that in general is different for each atom in a cluster, and sometimes even have different local easy axis directions. Furthermore, the 8 clusters considered in this figure all show rather different behaviors when it comes to the MAE. For some of them, e.g.,  site 3 in one cluster can have the $[100]$ easy axis direction, but other configurations could for this site favor the $[110]$ or the $[111]$ easy axis direction. 
As is clear from the figure we find values that are typically 5-10 times larger compared to the values of bcc Fe ($\sim 0.1 \mu$Ry) or fcc Ni ($\sim  0.2 \mu$Ry).~\cite{MAE-EXP} Further, these local MAE values are, remarkably, orders of magnitude larger compared to the almost vanishing value of the MAE of bulk Py. 

A key point in the Fig.~\ref{fig3} is that the symmetry of each cluster is not cubic. Hence, spin-orbit effects enter as a local uniaxial anisotropy and it depends of second-order anisotropy terms instead of fourth-order as cubic environments have. This is the primary reason why the local MAE values of Py are bigger than those of bcc Fe and fcc Ni.

It is interesting to compare the results of Fig.~\ref{fig2-new} and Fig.~\ref{fig3} to recent supercell calculations for the MAE of FeCo based alloys~\cite{Dane,Steiner}.  There it is also found that the local anisotropy of various atomic configurations varies strongly, and even changes sign, while the alloy MAE is described by the average over many configurations. Those systems are, however, very different in that they have a large MAE, meaning that one direction of magnetization should be over-represented among different cluster configurations. Py differs in that the MAE is vanishingly small, meaning that there must be a balance from different local anisotropy contributions.


\textit{Discussion and Conclusion --} 
We consider here the magnetic anisotropy of a macroscopic sample as a configurational average of local anisotropies, for a diverse distribution of clusters like the ones shown in Fig.~\ref{fig1}c. Each cluster may have several atoms with large local anisotropies directed in any of the common crystallographic axes ($\langle 001 \rangle$, $\langle 110 \rangle$ and $\langle 111 \rangle$), but since the inter-atomic exchange interaction of Py (not shown here) is much stronger and ferromagnetic, the resulting magnetic configuration is a collinear ferromagnet, where, after a proper configurational average is made, the resulting MAE is expected to be vanishingly small. 

In the presented study, were investigated only clusters with approximately the same concentration of Py ($\mathrm{Fe}_{0.2}\mathrm{Ni}_{0.8}$). In a real sample such constrain does not exist, and configurations involving, e.g., 1 Ni and 18 Fe atoms and vice-verse must also be considered. Once a proper configurational average of a huge set of clusters is considered, the  proper macroscopic MAE can be obtained, and we suggest this leads to a vanishingly small MAE for Py.  

The scenario proposed here is principally different than simply making a linear interpolation of anisotropy constants of bcc Fe and fcc Ni and adopting an interpolated value for all atoms of the alloy. We have shown that the local anisotropy is orders of magnitude larger than the observed anisotropy in Py. We therefore argue that the vanishing anisotropy in bulk Py arises from the cancellation of these local anisotropies. It is likely that the scenario put forward here also applies to other magnetic parameters, like the damping parameter or potentially the asymmetric exchange (like a local Dzyaloshinskii-Moriya interaction). We also note that experiments showed that amorphous materials present orbital induced magnetic anisotropy~\cite{Hase,Magnus} explained by the random anisotropy model. Note that in amorphous materials the lack of symmetry (chemical and crystalline) allows the emergence of orbital anisotropy. 

As a final comment, we note that the local anisotropy effects discussed here might affect the magnetization dynamics in thin films of Py~\cite{mauri}. For that, adopting a scenario of locally unique information, as proposed here, would be relevant for the interpretation of pump-probe measurements and crucial to simulations involving an effective spin-Hamiltonian.

\textit{Acknowledgments --} D.C.M.R thanks to D. Thonig for insightful discussions. The authors acknowledge support from the Swedish Research Council (VR), the KAW foundation (grants 2012.0031 and 2013.0020), STANDUPP and eSSENCE. A.B.K acknowledges support from CNPq (Brazil). D.C.M.R acknowledges support from CAPES (Brazil) for financial support and ``CENAPAD/UNICAMP" for providing computational facilities.

    \bibliographystyle{aipnum4-1.bst}
    \bibliography{References}






\end{document}